\documentclass{appolb}
\usepackage{epsfig}
\newcommand{\be}{\begin{equation}}
\newcommand{\ee}{\end{equation}}
\def\rs{\mbox{$\sqrt{s}$}}

\newcommand{\PR}[3]{{ Phys. Rev.}        {\bf #1} {(19#2)} {#3}}

\newcommand{\NP}[3]{{ Nucl. Phys.}       {\bf #1} {(19#2)} {#3}}
\begin{document}
\title{Total cross sections and soft gluon resummation%
\thanks{Presented at PLC2005, Kazimierz 5-8.09.2005}%
}
\author{R. M. Godbole
\address{Center for High Energy Physics, Indian Institute of Science, 
Bangalore,  560 012, India}
\and
A. Grau 
\address{Departamento de Fisica Teorica y del Cosmos, Universidad de 
Granada, Spain}
\and
G. Pancheri
\address{INFN, Laboratori Nazionali di Frascati, P.O. Box 13,
I-00044 Frascati, Italy}
\and 
Y.N. Srivastava
\address{Physics Department and INFN, University of Perugia, 
Perugia, Italy}
}
\maketitle
\begin{abstract}
We discuss a model for total hadronic and photonic cross-sections which
includes hard parton-parton scattering to drive the rise and soft gluon 
resummation to tame it. Unitarity is ensured by embedding the cross-section
in the eikonal formalism. Predictions for LHC and ILC are presented.
\end{abstract}
  
\section{Introduction}
Over three decades ago, the unexpected rise of the total proton-proton 
cross-section gave the first strong indications regarding  
the existence of hard scattering amongst the parton constituents of the 
proton 
\cite{therise}. And in the very near future, at LHC, a new much higher 
energy window for parton parton scattering will open up for which a precise 
knowledge of total cross-section predictions will be necessary to 
disentangle background from other processes and perhaps detect new physics. 
In this note we shall give an overview of the physical role of the various 
parameters entering most 
phenomenological descriptions of the total cross-section and subsequently 
present a model -which incorporates them- that can be used to predict the 
total cross-section at LHC and at the Photon Collider.

\section{Understanding the parameters in total cross-section models }

%\subsection{Subsection}
The proton  data exhibit, and require explanation of, three basic features: 
\begin{description}
\item (i) the normalization of the cross section, 
\item (ii) an initial decrease and 
\item (iii) a  subsequent rise with energy
\end{description}

Many models are available in the literature regarding the above issues, 
their predictions depending upon a number of parameters which are usually 
fixed by comparison with the low energy data. Before discussing some of 
these models and their predictions, we  shall provide phenomenomenological 
reasons for the approximate values of the parameters which are responsible 
towards a satisfactory description of the above  behaviours (i-iii).

 Let us recall that the decrease and the subsequent rise are well
  understood within a number of models as due to the exchange of a Regge
  and a Pomeron trajectory, through the expression \cite{DL}
\begin{equation}
\sigma_{tot}(s)=Xs^\epsilon+Ys^{-\eta}.
\label{DL}
\end{equation}
 The two terms of Eq.(\ref{DL}) reflect the
  well known duality between resonance and Regge pole exchange on the one hand
  and background and Pomeron exchange on the other, established in the late
  60's
  through FESR \cite{FESR}. This correspondance  meant that, while at 
  low energy the cross-section could be written as due to a  background term
  and a sum of resonances,   at higher energy it could be written as 
a sum of Regge trajectory exchanges and   a Pomeron exchange.  

  Before entering into the phenomenological analysis,
it is well to ask (i) where the  ``two component'' structure 
of Eq.(\ref{DL}) comes
from and (ii) why the difference in the two powers (in $s$) is 
approximately a half.

Our present knowledge of QCD and its employment for a description of
  hadronic phenomena  can and does provide some insight into the nature of 
these two terms.
%  and we  can then use quark hadron
%  duality, to see how at low and intermediate energies the properties of 
%quarks  and gluons manifest themselves as resonances and background 
%or as Regge
%  and Pomeron exchange.}
We shall begin answering the above two questions through  considerations 
about the 
bound state nature of hadrons which necessarily transcends perturbative 
QCD. For hadrons made of light quarks($q$) and glue($g$), the two terms 
arise from $q\bar{q}$ and $gg$ excitations. For these, the energy is given 
by a sum of three terms: (i) the rotational energy, (ii) the Coulomb 
energy and (iii) the ``confining'' energy. If we accept the Wilson area 
conjecture in QCD, (iii) reduces to the linear potential\cite{Widom1,
Widom2}. 
Explicitly, in the CM frame of two massless particles,
either a $q\bar{q}$ or a $gg$ pair separated by a relative 
distance $r$ with relative angular momentum $J$, the energy is given by
\begin{equation} \label{string1}
E_i(J, r) = {{2J}\over{r}} - {{C_i \bar{\alpha}}\over{r}} + C_i \tau r,
\end{equation}
where $i\ =\ 1$ refers to $q\bar{q}$,  $i\ =\ 2$ refers to $gg$, $\tau$
is the ``string tension'' and the Casimir's are $C_1\ =\ C_F\ =\ 4/3$, $C_2\
=\ C_G\ =\ 3$. $\bar{\alpha}$ is the QCD coupling constant evaluated at 
some average value of $r$ and whose precise  value
will disappear in the ratio to be considered. The hadronic
rest mass for a state of angular momentum $J$ is then computed through
minimizing the above energy
\begin{equation} \label{string2}
M_i(J) = Min_r[ {{2J}\over{r}} - {{C_i \bar{\alpha}}\over{r}} + C_i \tau r ],
\end{equation}
which gives
\begin{equation} \label{string3}
M_i(J) = 2 \sqrt{(C_i \tau)[2J - C_i \bar{\alpha}]}.
\end{equation}
The result may then be inverted to obtain the two sets of linear Regge 
trajectories $\alpha_i(s)$ 
%(which is {\it not} the coupling constant)
\begin{equation} \label{string4}
\alpha_i(s) = {{C_i \bar{\alpha}}\over{2}} + ({{1}\over{8 C_i \tau}}) s
= \alpha_i(0) + \alpha_i' s.
\end{equation} 
Thus, the ratio of the intercepts is given by
\begin{equation}\label{string5}
{{\alpha_{gg}(0)}\over{\alpha_{q\bar{q}}(0)}} = C_G/C_F = {{9}\over{4}}. 
\end{equation}
Employing our present understanding that resonances are $q{\bar q}$ bound 
states while the background, dual to the Pomeron,
 is provided by gluon-gluon exchanges\cite{landshoff}, the above 
equation can be rewritten  as 
\begin{equation} \label{string5qhd}
{{\alpha_{P}(0)}\over{\alpha_R(0)}} = C_G/C_F = {{9}\over{4}}. 
\end{equation}
If we restrict our attention to the leading Regge trajectory, namely
the degenerate $\rho-\omega-\phi$ trajectory, then
 $\alpha_R(0)=\eta\ \approx\ 0.48-0.5$, and we obtain
for $\epsilon\ \approx\ 0.08-0.12$, a rather  satisfactory value. 
The same argument for the slopes gives 
\begin{equation}\label{string6}
{{\alpha_{gg}'}\over{\alpha_{q\bar{q}}'}} = C_F/C_G = {{4}\over{9}}.
\end{equation}
so that if we take for the Regge slope $\alpha_R'\ \approx\ 0.88-0.90$, we get
for $\alpha_P'\ \approx\ 0.39-0.40$,  in fair
agreement with lattice estimates\cite{lattice}.

%We are not aware of any alternative explanation for these facts: neither 
%for the need of two components nor for the ratio of the two intercepts.   

We now have good reasons for a break up of the amplitude into 
two components. To proceed further, it is necessary to realize that 
precisely because massless hadrons do not exist,  Eq.(\ref{DL}) violates 
the Froissart bound and thus must be unitarized. To begin this task, 
let us first rewrite  Eq.(\ref{DL}) by putting in the  ``correct'' dimensions
\begin{equation} \label{DL1} 
\bar{\sigma}_{tot}(s)= \sigma_1 (s/\bar{s})^\epsilon+ \sigma_2
(\bar{s}/s)^{1/2},
\end{equation}
where we have imposed the nominal value $\eta\ =\ 1/2$. In the following,
we shall obtain rough estimates for the size of the parameters in 
Eq.(\ref{DL1}). 

A minimum occurs
in $\bar{\sigma}_{tot}(s)$ at $s\ =\ \bar{s}$, for $\sigma_2\ =\ 2\epsilon
\sigma_1$.
%, so that
If we make this choice, then Eq.(\ref{DL1}) has one less parameter 
and it reduces to
\begin{equation} \label{DL2} 
\bar{\sigma}_{tot}(s)= \sigma_1 [(s/\bar{s})^\epsilon+ 2 \epsilon
(\bar{s}/s)^{1/2}].
\end{equation}
We can isolate the rising part of the cross-section by rewriting the above
as
\begin{equation}\label{DL3}
\bar{\sigma}_{tot}(s)= \sigma_1 [ 1 + 2\epsilon(\bar{s}/s)^{1/2}] 
+ \sigma_1 [(s/\bar{s})^\epsilon - 1].
\end{equation}
Eq.(\ref{DL3}) separates cleanly the cross-section into two parts: the first
part is a ``soft'' piece which shows a saturation
to a constant value (but which contains no rise) and the second a ``hard'' 
piece which has all the rise. Morover, $\bar{s}$ naturally provides the 
scale beyond which the cross-sections would begin to rise. Thus, our
``Born''  term assumes the generic form
\begin{equation} 
\label{DL4}
\sigma_{tot}^B(s)= \sigma_{soft}(s) + \vartheta (s - \bar{s})
\sigma_{hard}(s).
\end{equation}
with $\sigma_{soft}$ containing a constant ( the ``old'' Pomeron
with $\alpha_P(0)\ =\ 1$) plus a (Regge) term decreasing
as $1/\sqrt{s}$ and with an estimate for their relative magnitudes
($\sigma_2/\sigma_1\ \sim\ 2\epsilon$). We shall assume that the rising 
part of the cross-section $\sigma_{hard}$ is provided by jets which are 
calculable by perturbative QCD, obviating (atleast in principle) the 
need of an arbitrary parameter $\epsilon$.

       An estimate of $\sigma_1$ may also be obtained through the hadronic
string picture. Eq.(\ref{string2}) gives us the mean distance between quarks
or the ``size'' of a hadronic excitation of angular momentum $J$ in 
terms of the string tension
\begin{equation} 
\label{string7}
\bar{r}(J)^2 = {{2J - C_F\bar{\alpha}}\over{\tau}}.
\end{equation}

Thus, the size $R_1$ of the lowest hadron (which in this Regge string
picture has $J\ =\ 1$, since $\alpha_R(0)\ =\ 1/2$) is given by
\begin{equation} \label{string8}
R_1^2 = {{1}\over{\tau}} = 8 \alpha'
\end{equation} 
If two hadrons each of size $R_1$ collide, their effective radius
for scattering would be given by
\begin{equation} \label{string9}
R_{eff} = \sqrt{R_1^2 + R_1^2} = \sqrt{2} R_1,
 \end{equation}    
and the constant cross-section may be estimated (semi-classically) to be roughly
\begin{equation}
 \label{string10}
\sigma_1 = 2 \pi R_{eff}^2 = 4 \pi R_1^2 \approx {{4\pi}\over{\tau}}
= 32 \pi \alpha', 
\end{equation}
which is about $40\ mb$, a reasonable value. 
In the later sections, for the ``soft'' cross-section we shall take a value
        of this order of magnitude  as the nominal value.
  
The last remaining parameter is the scale $\bar{s}$, the 
jet production threshold in the hadronic cross-section. In $e^+e^-$
annihilation, the threshold for jet production can be estimated from the
appearance of multihadronic production in $e^+e^-$ scattering first
observed at ADONE  around $3$ GeV.
For scattering of two hadrons, this should translate into an
$\sqrt{\bar{s}}\ \approx\ 12 $ GeV. Thus, from Eq.(\ref{DL3}), we have
\begin{equation}
 \sigma_1 [ 1 + 2\epsilon(\bar{s}/s)^{1/2}] \approx \sigma_1 
(1+{{2}\over{\sqrt{s}}})  
\end{equation}
The above phenomenological estimate holds for proton-antiproton scattering,
whereas for proton-proton, which has no resonances in the s-channel, no
Regge exchange is expected to contribute and  only the (approximately) 
constant term remains.
\section{The Bloch-Nordsieck model for LHC}

In the past, several authors realized that QCD offers an elegant explanation of 
the rise of the cross-section via the minijets and hence suggested that the rise 
of $\sigma_{tot}$ with energy was driven by the rapid rise with energy of the 
inclusive jet cross-section  
\be \label{sigjet}
\sigma^{ab}_{\rm jet} (s) = \int_{p_{tmin}}^{\rs/2} d p_t
\int_{4 p_t^2/s}^1 d x_1 \int_{4 p_t^2/(x_1 s)}^1 d x_2 \sum_{i,j,k,l}
f_{i|a}(x_1) f_{j|b}(x_2) \frac { d \hat{\sigma}_{ij \rightarrow kl}(\hat{s})}
{d p_t},
\ee
In an eikonal minijet model (EMM), the total cross-section then reads
\begin{equation}
\sigma_{tot}\approx 2 \int d^2{\vec b}[1-e^{-n(b,s)/2}]
\end{equation}
wherein
\begin{equation}
n(b,s)=2\Im
\chi(b,s)= n_{soft}+n_{hard}=A_{soft}(b)\sigma_{soft}(s)+A_{jet}(b)\sigma_{jet}(s)
\label{nsplit}
\end{equation}
and $\Re \chi(b,s)=0$. In the Bloch-Nordsieck (BN) model \cite{ff2}, 
the overlap functions $A_i(b)$ are s-dependent and given by 
\begin{equation}
\label{abn}
A_{BN}={{e^{-h(b,s)}}\over{\int d^2{\vec b}
e^{-h(b,s)}}},
\end{equation}
\begin{equation}
\label{hb}
h(b,s)={{8}\over{3\pi}}\int_0^{q_{max}}{{dk}
\over{k}}
\alpha_s(k^2)\ln({{q_{max}+\sqrt{q_{max}^2-k^2}}
\over{q_{max}-\sqrt{q_{max}^2-k^2}}})[1-J_0(kb)]
\end{equation}
and  $q_{max}$ 
%is a function of energy, which
depends on energy and the kinematics of the process\cite{greco}.

%\begin{equation}
%h(b)=b^2p_{\perp int}^2 + S(b)
%\end{equation}
%where the intrinsic transverse momentum $p_{\perp int}$ is a constant, of the 
%order of a few 100 MeV, parametrized according to the process under 
%consideration. 
The eikonal formalism which we use to describe the total
cross-section, incorporates
multiple parton parton collisions, accompanied by
soft gluon emission from the initial valence
quarks, to leading order. Notice that in this
model, we consider emissions only from the
external quark legs. In  the impulse approximation -on 
which the parton model itself is based-  the valence quarks 
are free, external particles. In this picture, emission of 
soft gluons from the gluons involved in the  hard  scattering, 
is non leading. As the energy increases, more and more hard
gluons are emitted but  there is also a larger and larger 
probability of soft gluon emission : the overall effect is a 
rise of the cross-section, tempered  by the soft emission, 
i.e. the violent mini-jet rise due to semi-hard gluon gluon 
collisions is tamed by soft gluons. Crucial in this model, are 
the scale and the behaviour of the strong coupling constant 
which is present in the integral over the soft gluon spectrum. 
While in the jet cross-section $\alpha_s$ never plunges into
the infrared region, as the scattering partons are by 
construction semi-hard, in the soft gluon spectrum, the opposite 
is true and a regularization is mandatory. We notice however 
that here, as in other problems of soft hadron physics\cite{doksh}, 
what matters most is not the value of $\alpha_s(0)$, but rather 
its integral.  Thus, all that we need to demand, is that $\alpha_s$ 
be integrable, even if singular~\cite{nak}. We employ the same 
phenomenological expression for $\alpha_s$ as used in our 
previous works, namely
\begin{equation}
\label{alphaRich}
\alpha_s(k_\perp)={{12 \pi }\over{(33-2N_f)}}{{p}\over{\ln[1+p({{k_\perp}
\over{\Lambda}})^{2p}]}}
\end{equation}
Through the above, we were able to reproduce the effect of the
phenomenologically introduced intrinsic transverse momentum of 
hadrons~\cite{nak}, and more recently obtained a very good 
description of the entire region where the total cross-section 
rises \cite{ff2}. 
This expression for $\alpha_s$ coincides with the usual one-loop expression 
for large values of $k_\perp$, while going to a singular limit for small 
$k_\perp$. For $p=1$ this expression corresponds to the Richardson
potential\cite{richardson} used in bound state problems. We see 
from Eq.(\ref{hb}) that $p\ =\ 1$, leads to a divergent integral, and thus 
cannot be used. Notice that, presently, in the  expression for $h(b,s)$, 
the masses of the emitting particles are put to zero as is
usual in perturbative QCD. Thus, for a convergent integral,
 one requires $p<1$ and the successful 
phenomenology indicated in \cite{ff2} gave $p=3/4$.
%The QCD running coupling constant has been parametrized as follows
%to account for asymptotic freedom as well as confinement:
%\begin{equation}
%\label{alphaRich}
%\alpha_s(k_\perp)={{12 \pi }\over{(33-2N_f)}}{{p}\over{\ln[1+p({{k_\perp}
%\over{\Lambda}})^{2p}]}}
%\end{equation}

\begin{figure}[htbp!]
\begin{center}
\label{lhc}
\epsfig{file=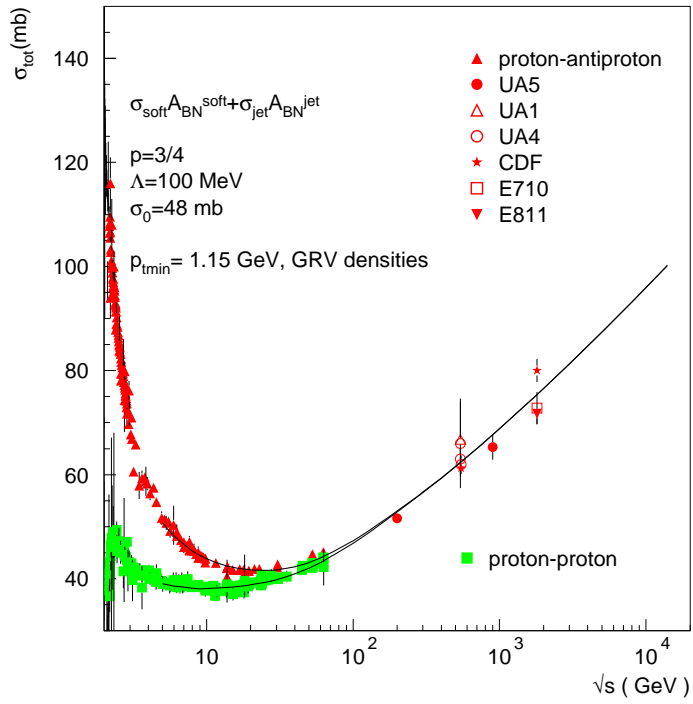, width=10cm}
\caption{Comparison of $p p$ and $p\bar{p}$ 
total cross-section data \cite{PDG05,UA4,UA1,UA5,E710,CDF,E811} with 
predictions for the total $pp$ and $p{\bar p}$ cross-section from the QCD model described in the text for an optimal choice of parameters. }
\end{center}
\end{figure}
\section{The photon-photon total cross-section}
We now show an application of the above model\cite{albert} to photon photon scattering
and its comparison with present data \cite{L3,OPAL} and with another 
model\cite{block}.
\begin{figure}[htbp!]
\begin{center}
\label{plc}
\epsfig{file=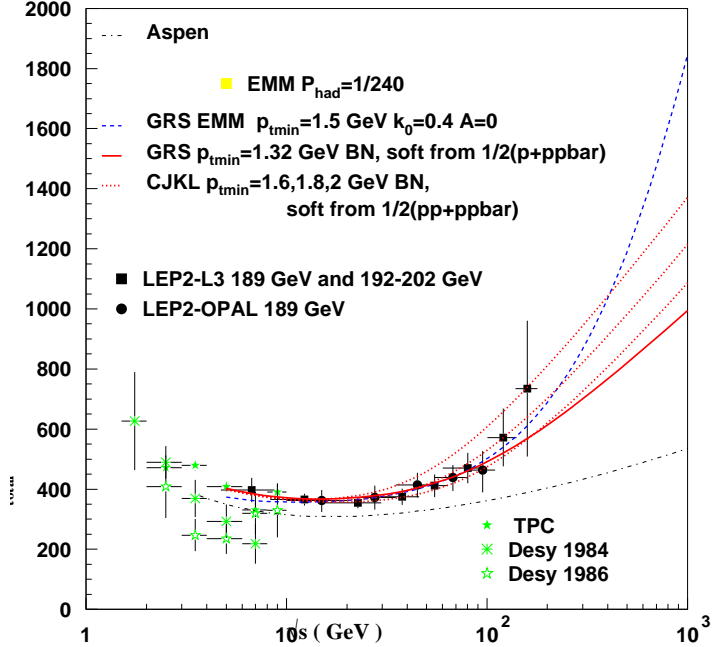, width=10cm}
\vskip 1cm
\caption{Predictions for the total photon-photon cross-section based on the
QCD  model described in the text and a parameter set consistent with those 
for proton and 
proton-antiproton scattering}
\end{center}
\end{figure}
\section*{Ackowledgments}
This work was supported in part through EU RTN Contract CT2002-0311.
RG wishes to acknowledge the partial support of the the Department of 
Science and Technology, India, under project number SP/S2/K-01/2000-II.
AG acknowledges support from MCYT under project number FPA2003-09298-c02-01.


\begin{thebibliography}{99}
\bibitem{therise}
D. Cline, F. Halzen and J. Luthe,  Phys. Rev. Lett. {\bf 31} (1973)
491.
\bibitem{DL}A. Donnachie and P.V. Landshoff, {\it Phys. Lett.}
{\bf B296}(1992)227.
\bibitem{FESR} K. Igi and S. Matsuda, {\it Phys Rev Lett} 
{\bf 18} (1967) 625.
\bibitem{Widom1} 
Y. Srivastava, S. Pacetti, G. Pancheri, A. Widom, Invited talk at
``$e^+e^-$ Physics at Intermediate Energies Workshop'';
ArXiv: hep-ph/0106005; Journal-ref: eConf C010430 (2001) T19.
\bibitem{Widom2}
Y. Srivastava and A. Widom, {\it Phys.Rev.} {\bf D63} (2001) 077502.
\bibitem{landshoff}
P. V. Landshoff, ArXiv:hep-ph/0108156.
\bibitem{lattice}
H. B. Mayer and M. J. Teper, {\it Phys Lett} {\bf B605} (2005) 344.
\bibitem{ff2}A. Grau, G. Pancheri and Y.N. Srivastava,
Phys.Rev. {\bf D60} (1999) 114020; R.M. Godbole, A. Grau, G. Pancheri and
Y.N. Srivastava, Phys. Rev. {\bf D72} (2005) 076001.
\bibitem{greco}P. Chiappetta and M. Greco, \NP{B199}{82}{77}.
\bibitem{doksh}
Y.L. Dokshitzer, 
{\it Perturbative QCD
 Theory (includes our knowledge of $\alpha_s)$}, Proceedings of 
ICHEP 1998, Vancouver, hep-ph/9812252.
\bibitem{nak}
A. Nakamura, 
G. Pancheri and Y. Srivastava, Z.\ Phys.\ {\bf C21} (1984) 243.
\bibitem{richardson}
J.L. Richardson, Phys.\ Lett.\ {\bf B82} (1979) 272.
\bibitem{PDG05} S. Eidelman {it et al.}, Phys.\ Lett.\ {\bf B592} (2004) 1.
\bibitem{UA4} UA4 Coll. (R. Battiston et al.), 
Phys.\ Lett.\ {\bf 117B} (1982) 126;
UA4 Coll. (M. Bozzo et al.), Phys.\ Lett.\ {\bf 147B} (1984) 392;
UA4/2 Coll. (C. Augier et al.), Phys.\ Lett.\ {\bf 344B} (1994) 451.
\bibitem{UA1}UA1 Coll. (Arnison et al.), Phys.\ Lett.\ {\bf 128B} (1983) 336.
\bibitem{UA5}UA5 Coll. (G. J. Alner et al.), Zeit.\ Phys.\ {\bf C32} 
(1986) 153.
\bibitem{E710} E710 Coll. (N. Amos et al.), 
Phys.\ Rev.\ Lett.\ {\bf 68} (1992) 2433.
\bibitem{CDF}CDF Coll. (F. Abe et al.), Phys.\ Rev.\ {\bf D50} (1994) 5550.
\bibitem{E811}E811 Coll. (C. Avila et al.), Phys.\ Lett.\ {\bf 445B}
  (1999).
\bibitem{albert}A. de Roeck, R.M. Godbole, A. Grau, G. Pancheri and
  Y.N. Srivastava, {\it Total cross-sections: Cross talk between HERA, LHC 
and LC}, LCWS 2004 Proceedings, Paris April 2004, Eds. H. Videau and
J-C. Brient, Editions de l'Ecole Politechnique-juillet 2005; hep-ph/0412189.
\bibitem{L3} L3 Collaboration, M. Acciarri et al.,  
Phys. Lett. {\bf B 408} (1997)  450; Phys.Lett. {\bf B519} (2001) 33, 
hep-ex/0102025. 
%Paper 519 submitted to {\it ICHEP'98}, Vancouver, July 1998;
L3 Collaboration, A. Csilling,  Nucl. Phys. Proc. Suppl. {\bf B82} 
(2000) 239.
%; L3 Note 2548, Submitted to the {\it OSAKA Conference}. 
\bibitem{OPAL}OPAL Collaboration. 
G. Abbiendi et al., Eur. Phys. J.  {\bf C14} (2000)  199;  
F. Waeckerle,  Nucl. Phys. Proc. Suppl. {\bf B71} (1999) 381; 
Stefan S\"oldner-Rembold, hep-ex/9810011.
\bibitem{block} M.M. Block, F. Halzen, B. Margolis, \PR{D45}{92}{839};
M. Block, E. Gregores, F. Halzen and G. Pancheri, \PR{D58}
{98}{17503}.
\end{thebibliography}
\end{document}